\documentclass[pdftex,preprint2]{aastex61}
\usepackage{apjfonts}
\usepackage{color}
\definecolor{20170927}{rgb}{0,0,0}
\definecolor{20180301}{rgb}{0,0,0}
\received{\today}
\revised{\today}
\accepted{\today}
\submitjournal{ApJ}
\shorttitle{MIR ``Slow-Scanning'' Observations}
\shortauthors{Ohsawa et al.}
\begin{document}
\title{``Slow-Scanning'' in Ground-Based Mid-Infrared Observation\textcolor{20170927}{s}}
\correspondingauthor{Ryou Ohsawa}
\email{ohsawa@astron.s.u-tokyo.ac.jp}
\author{Ryou Ohsawa}
\affiliation{Institute of Astronomy, Graduate School of Science, University of Tokyo, 2-21-1 Osawa, Mitaka, Tokyo 181-0015, Japan}
\author{Shigeyuki Sako}
\affiliation{Institute of Astronomy, Graduate School of Science, University of Tokyo, 2-21-1 Osawa, Mitaka, Tokyo 181-0015, Japan}
\affiliation{Precursory Research for Embryonic Science and Technology (PRESTO), Japan Science and Technology Agency (JST), 2-21-1 Osawa, Mitaka, Tokyo, 181-0015, Japan}
\author{Takashi Miyata}
\author{Takafumi Kamizuka}
\author{Kazushi Okada}
\author{Kiyoshi Mori}
\author{Masahito S. Uchiyama}
\author{Junpei Yamaguchi}
\affiliation{Institute of Astronomy, Graduate School of Science, University of Tokyo, 2-21-1 Osawa, Mitaka, Tokyo 181-0015, Japan}
\author{Takuya Fujiyoshi}
\affiliation{Subaru Telescope, National Astronomical Observatory of Japan, National Institutes of Natural Sciences, 650 North A'ohoku Place, Hilo, HI 96720, USA}
\author{Mikio Morii}
\author{Shiro Ikeda}
\affiliation{The Institute of Statistical Mathematics, 10-3 Midori-cho, Tachikawa, Tokyo 190-8562, Japan}
\begin{abstract}
Chopping observations with a tip-tilt secondary mirror have conventionally been used in ground-based mid-infrared observations. However, it is not practical for next generation large telescopes to have a large tip-tilt mirror that moves at a frequency larger than a few Hz. We propose an alternative observing method, a ``slow-scanning'' observation. Images are continuously captured as movie data, while the field-of-view is slowly moved. The signal from an astronomical object is extracted from the movie data by a low-rank and sparse matrix decomposition. The performance of the ``slow-scanning'' observation was tested in an experimental observation with \textit{Subaru}/COMICS. The quality of a resultant image in the ``slow-scanning'' observation was as good as in a conventional chopping observation \textcolor{20180301}{with COMICS}, at least for a bright point-source object. The observational efficiency in the ``slow-scanning'' observation was better than that in the chopping observation. The results suggest that the ``slow-scanning'' observation \textcolor{20180301}{can be a competitive method for the \textit{Subaru} telescope and be of potential interest to other ground-based facilities to avoid chopping}.
\end{abstract}
\keywords{%
methods: data analysis ---
methods: observational ---
techniques: image processing
}
\section{Introduction}
\label{sec:intro}
Observations in the mid-infrared have been widely used to trace warm interstellar matter. Both spectroscopy and imaging observations are useful to investigate, for example, circumstellar envelopes of evolved stars, evolution of micrometer dust grains in protoplanetary disks, and star-forming activity in massive star-forming regions.
Mid-infrared observations have been carried out with ground-based telescopes despite heavy atmospheric absorption. Large ground-based telescopes enable observations at high resolution: with a typical seeing size of ${\sim}0.3''$ for $10\,$m class telescopes without adaptive optics \citep{kataza_comics:_2000}. While there are several mid-infrared space observatory projects ongoing, ground-based mid-infrared observations will remain important for studies requiring a high-spatial resolution \citep[e.g.,][]{ohsawa_unusual_2012,honda_high-resolution_2015,matsumoto_mid-infrared_2008,miyata_sub-arcsecond_2004}.
A high-frequency beam switching technique (hereafter chopping) has been widely used in ground-based mid-infrared observations \textcolor{20170927}{\citep[e.g.,][]{allen_infrared:_1975,papoular_processing_1983,beckers_imaging_1994}}. A large fraction of the signal originates from terrestrial atmosphere, the telescope, and the instrument itself. The atmospheric emission rapidly changes, like a $1/f$-noise \textcolor{20170927}{\citep{kaeufl_sky-noise_1991,westphal_infrared_1974,bensammar_statistical_1978,allen_study_1981}}, indicating that the noise from atmospheric emission cannot be eliminated just by integration. A pixel-to-pixel response map (widely referred to as a flat frame) is hardly well-defined. Thus, it is practically impossible to estimate the intensity of background emission from adjacent regions. \textcolor{20170927}{A dithering observation, typically used in near-infrared observations, does not work in the mid-infrared.}
Chopping observations are one solution to these difficulties. By differentiating chop pairs, the atmospheric emission can be greatly suppressed. A chop pair of images should be taken before the atmospheric emission varies. \textcolor{20170927}{The required chopping frequency depends on the wavelength. \citet{kaeufl_sky-noise_1991} suggested that the chopping frequency should be higher than about $8\,\mathrm{Hz}$ at $10\,\mathrm{\mu m}$.}
A tip-tilt secondary mirror is widely used for chopping observations \citep[e.g.,][]{lorell_high_1993}. By oscillating the secondary mirror at about a few Hz, the beam is quickly switched. Thus, ground-based telescopes without such oscillating secondary mirrors are currently not capable of mid-infrared observations. The situation is worse for next-generation $30\,$m-class telescopes: their secondary mirrors are too large to be oscillated at high frequencies such as $\gtrsim 1\,$Hz.
Recently, a cold chopper has been proposed instead of the tip-tilt secondary mirror. By installing a tip-tilt mirror with a cold optics, quick beam switching is achievable. \citet{paalvast_development_2014} developed a cryogenic tip-tilt mirror for the METIS \citep{brandl_status_2016,brandl_metis:_2014}. MAX38, a mid-infrared camera mounted on the miniTAO 1.0-m telescope \citep{minezaki_university_2010}, successfully obtained scientific images using its cold chopper \citep{nakamura_minitao/max38_2010}. The technology required for a cold chopper is, however, under development. It is not easy to achieve a long stroke, fast movement, and small heat production at the same time \citep{mori_development_2016}.
Other observational methods have been proposed. The ``drift scanning'' method is one of them \citep{heikamp_drift_2014}: images are continuously obtained at ${\gg}1\,$Hz while moving the telescope. This method is analogous to ``on-the-fly'' observations in radio astronomy \citep{sawada_--fly_2008,mangum_fly_2007}. The background emission is estimated using the frames taken before the atmospheric emission has changed. The ``drift scanning'' method does not require a tip-tilt secondary mirror. Instead, the telescope should be moved \textcolor{20170927}{before the atmospheric emission changes}. The ``weighted average'' method, which is similar to the LOCI algorithm \citep{marois_exoplanet_2010}, uses a sequence of images obtained in different positions to recompose the atmospheric emission in each pointing by a linear combination of the obtained images \citep{nakamura_method_2016}. A typical timescale for beam-switching in the ``weighted average'' method is about 15\,s. This method is also useful for reducing data obtained in chopping observations with insufficient frequency.
We propose another observational method for ground-based mid-infrared observations. In the proposed method, images are continuously obtained while the field-of-view is slowly shifted (hereafter, referred as to the ``slow-scanning'' method). The ``slow-scanning'' method does not require a tip-tilt secondary mirror nor an internal tip-tilt mirror. Neither fast telescope movement nor frequent beam-switching is required. We develop an algorithm to extract scientific information using redundancy in the data. The paper is organized as follows: Section~\ref{sec:algorithm} describes the formalism and the algorithm used in this paper. Details of the observations and results of the data reduction are described in Section~\ref{sec:observation}. The performance of the proposed method is discussed in Section~\ref{sec:discussion}. The paper is summarized in Section~\ref{sec:conclusion}.
\section{Method}
\label{sec:algorithm}
\subsection{``Slow-Scanning'' Observation}
\label{sec:slow-scanning}
\begin{figure}[tp]
\centering
\plotone{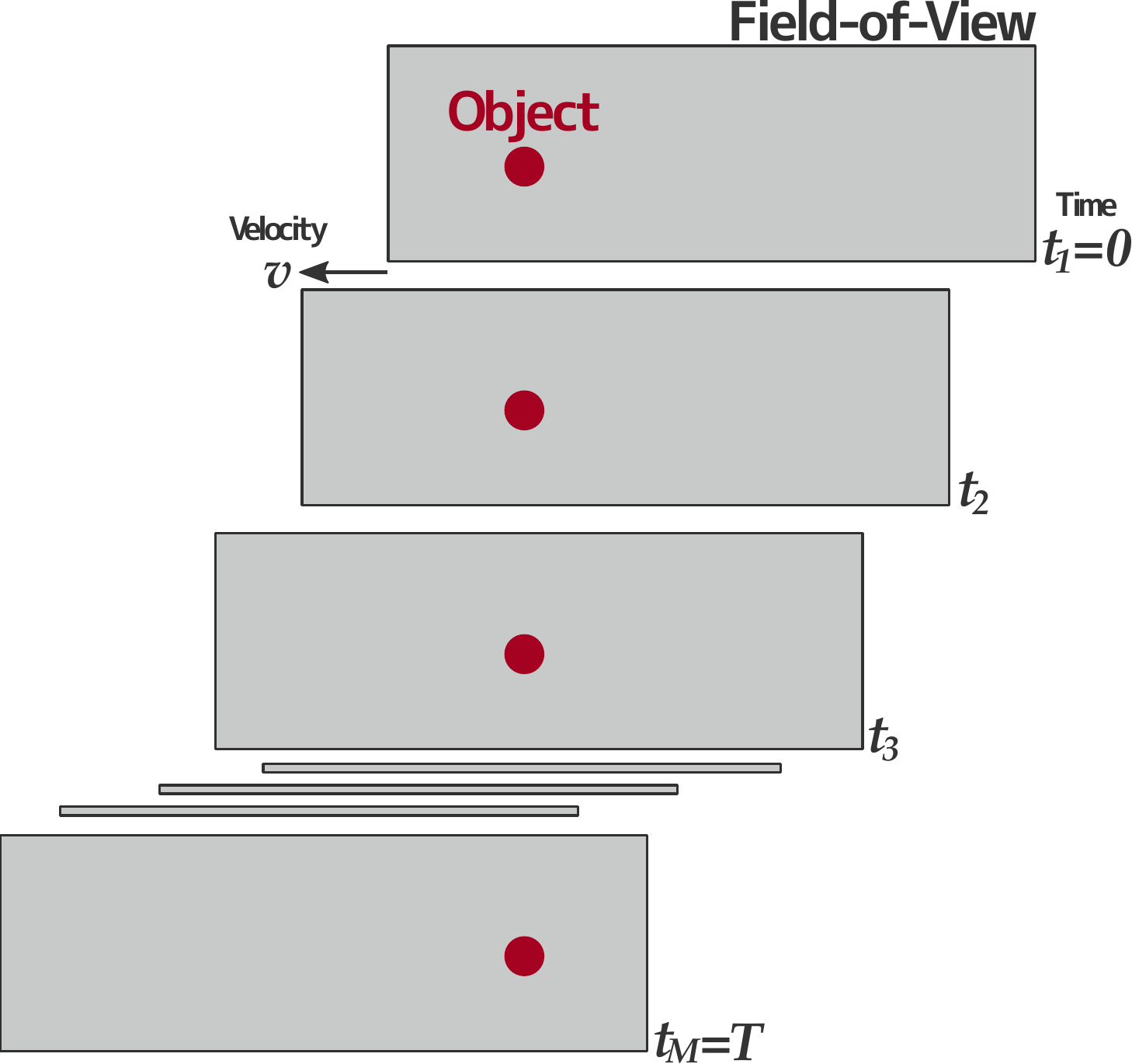}
\caption{Schematic view of the ``Slow-Scanning'' observation. The gray rectangles indicate the field of view of a single exposure. Images are continuously obtained by slightly shifting the field of view.}
\label{fig:scheme}
\end{figure}

The ``slow-scanning'' method is designed to extract scientific signals from a sequence of images. Figure~\ref{fig:scheme} shows a schematic view of the ``slow-scanning'' observation. In the observation, images are continuously obtained as movie data, while a telescope is slowly moved. The integration time of a single exposure should be shorter than the typical timescale of variation of atmospheric emission (${\lesssim}1\,$s). The scanning speed of the telescope should be slow enough, so that the image of the scientific target should not be affected by the telescope movement. For a point-source object, the scanning speed $v$ should be slower than ${\sim}{\theta}t^{-1}$, where $\theta$ is the pixel scale of a detector and $t$ is the integration time per exposure. The observation should be continued until the location of the target is entirely changed. The total observing time $T$ should be a few times larger than $\Theta {v^{-1}}$, where $\Theta$ is the angular size of the target. The movie data are compiled into a three-dimensional FITS data cube, which is the typical data structure in ground-based mid-infrared observations.
\subsection{Data Reduction Algorithm}
\label{sec:data-reduction}
\begin{figure*}[tp]
\centering
\plotone{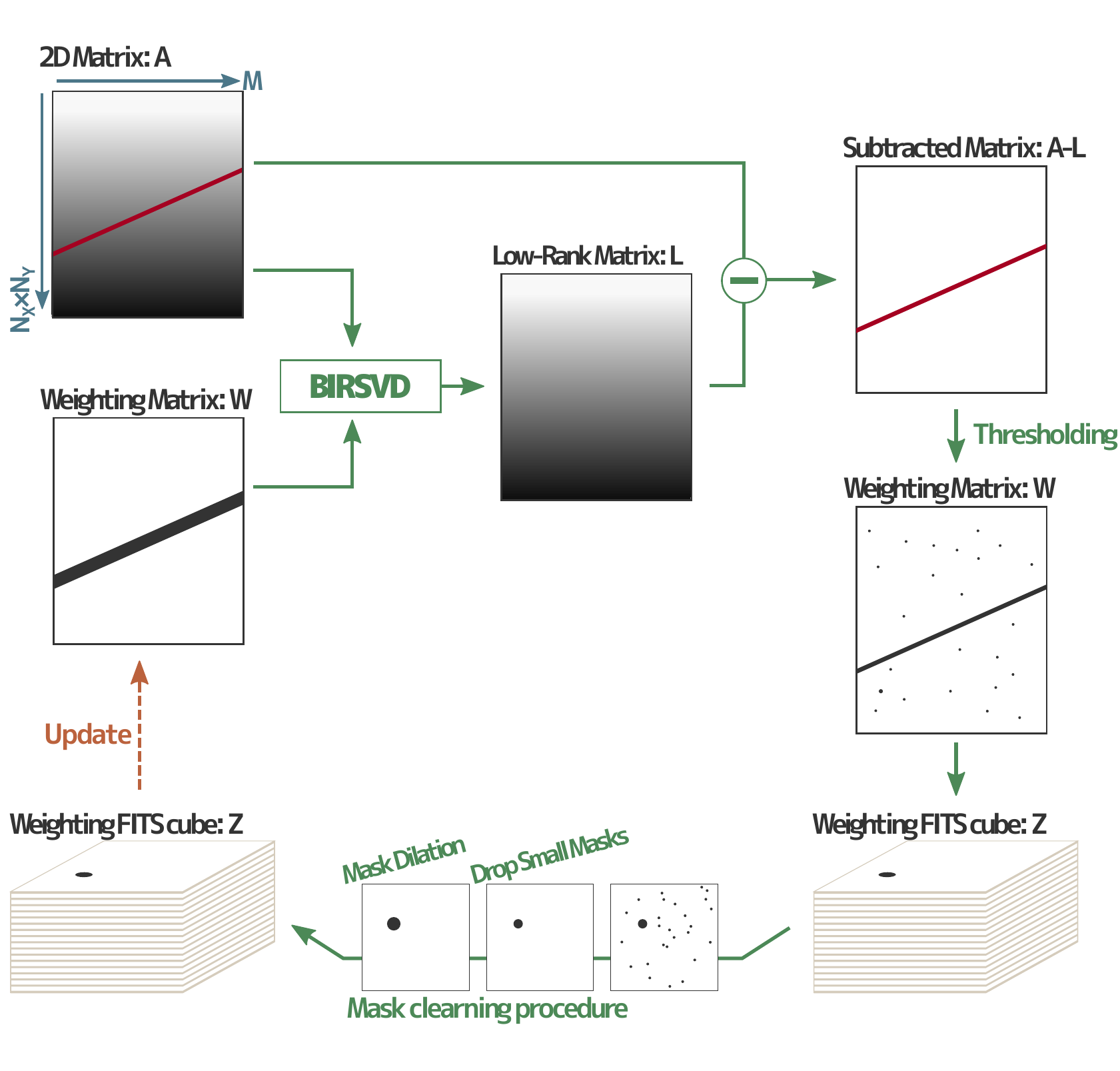}
\caption{Schematic view of the Slow-Scanning Decomposition Procedure.}
\label{fig:ssdec}
\end{figure*}

This section provides an algorithm to extract scientific information from the obtained data cube. First, characteristics of data obtained in the ground-based mid-infrared observation are described. An image with $N_X{\times}N_Y$ pixels is produced in every exposure. The position of a pixel is designated by an index $i$ and $j$. The images are continuously obtained: The total number of exposures is $M$. The time of the $m$th exposure is denoted by $t_m$. The signal at $(i,\,j)$ and the time $t_m$ is denoted by $I_{i,j,m} := I_{i,j}(t_m)$.
We assume that the signal in each pixel consists of emission from an astronomical object, atmospheric emission, emission from the telescope and instrument, and instrumental bias. The emission from the object (${\mathcal S}$) is assumed to be constant in time, but it changes its location with movement of the telescope. The atmospheric emission (${\mathcal A}$) is assumed to rapidly vary with time but marginally depends on $i$ and $j$. The emission from the telescope and the instrument (${\mathcal T}$) is expected to be constant in time while the temperature of the system remains the same. We assume that the bias, such as dark current, (${\mathcal D}$) is constant in time, but can vary with position in the image.
The signal originating from photons is subject to a pixel-to-pixel variation in the response of the detector, or the flat frame. \textcolor{20170927}{In general, the flat frame ($f$) is time-variant since the detector and a readout system are not stable. \citet{nakamura_method_2016}, however, successfully subtracted background emission with the ``weighted average'' method, where the stability of the flat frame was assumed. Thus, we expect that the flat frame is approximately stable over a time scale of a few minutes. For the sake of simplicity, we assume that $f$ is constant in time during the observation. The validity of this assumption is discussed in Section~\ref{sec:applicability} and Appendix~\ref{sec:flatframe}.} Thus, the signal value at $(i,\,j)$ in the $m$th exposure is described as
\begin{equation}
\label{eq:signal}
I_{i,j,m}
= f_{i,j}\bigl(
{\mathcal A}_{i,j,m}
+ {\mathcal T}_{i,j}
+ {\mathcal S}_{i,j,m}
\bigr) + {\mathcal D}_{i,j}.
\end{equation}
Statistical noise components (e.g., readout noise) are dropped here for the sake of simplicity. In a typical mid-infrared observation, the intensity of ${\mathcal S}$ is much lower than those of ${\mathcal A}$ and ${\mathcal T}$. A precise subtraction of ${\mathcal A}$ and ${\mathcal T}$ is required to extract the signal from astronomical objects. In $N$-band observations (${\sim}10\,\mu$m), ${\mathcal A}$ is as strong as ${\mathcal T}$, while ${\mathcal A}$ becomes much stronger at longer wavelengths.
Here, we describe how to extract the scientific signal of an astronomical object from the obtained data. The emission from the telescope and the instrument $f_{i,j}{\mathcal T}_{i,j}$ and the bias component ${\mathcal D}_{i,j}$ are assumed to be almost constant in time. The atmospheric component $f_{i,j}{\mathcal A}_{i,j,m}$ rapidly varies in time, but it is expected to be smooth in the spatial dimensions. These components are expected to have high redundancy in the obtained data. Instead, the scientific component $f_{i,j}{\mathcal S}_{i,j,m}$ contains the spatially-localized signal, which moves with time. This component is expected to have much less redundancy than the other components. Thus, an algorithm, which isolates non-redundant components from redundant components, will extract $f_{i,j}{\mathcal S}_{i,j,m}$ from the movie data obtained in the ``slow-scanning'' observation.
A low-rank and sparse matrix decomposition is widely used to extract non-redundant components. Several algorithms for \textcolor{20170927}{the matrix decomposition} have been developed \citep[e.g.,][]{candes_robust_2009,keshavan_gradient_2009,keshavan_matrix_2009}.
We focus on the Go Decomposition \citep[\texttt{GoDec};][]{zhou_godec:_2011}, which is a fast and robust implementation of the low-rank and sparse matrix decomposition. \texttt{GoDec} approximates the matrix $A$ by $L{+}S$, where $L$ is a low-rank matrix or a redundant component, and $S$ is a sparse matrix or a non-redundant component. The rank of the low-rank matrix $L$ is defined by $r$. The sparse matrix $S$ contains $k$ non-zero components. The matrices $L$ and $S$ are alternatively updated: $L$ is given by the low-rank matrix approximation of $A{-}S$, while $S$ consists of the non-zero subset of the first $k$ largest entries of $|A{-}L|$. \citet{morii_data_2017} applied the \texttt{GoDec} method to astronomical movie data and demonstrated that \texttt{GoDec} successfully extracts fast transient signals from background emission.
We propose a reduction procedure for the ``slow-scanning'' observation, which basically follows the procedure of \texttt{GoDec} but is slightly modified to work on masked data (\textit{hereafter}, referred as to masked GoDec or \texttt{mGoDec}). The procedure is described in Table~\ref{algo:maskedgodec}. Figure~\ref{fig:ssdec} illustrates a schematic view of the main iteration part in the \texttt{mGoDec} procedure. While the bilateral random projections method is implemented for the low-rank matrix approximation in \texttt{GoDec}, \texttt{mGoDec} uses Bi-Iterative Regularized Singular Value Decomposition \citep[\texttt{BIRSVD};][]{das_fast_2011,das_birsvd_2011}\footnote{The \texttt{BIRSVD} software is available at \url{http://www.mat.univie.ac.at/~neum/software/birsvd/}}.
\textcolor{20170927}{First, the obtained data $I_{i,j,m}$, a three-dimensional data cube $\left(\mathbb{R}^{N_X} {\times} \mathbb{R}^{N_Y} {\times} \mathbb{R}^M\right)$, is rearranged into a two-dimensional matrix $A_{n,m}\,\left(\mathbb{R}^{N_XN_Y} {\times} \mathbb{R}^M\right)$  (Line~1--3 in Table~\ref{algo:maskedgodec}).} The \texttt{BIRSVD} method generates a low-rank approximation of the matrix weighted by a weighting matrix $W$ (Line~5). The weighting matrix $W$ is a binary mask matrix such that $W_{n,m} = 1$ when the $(n,\,m)$-element of a matrix is masked. The weighting matrix $W$ is updated by ``thresholding'' (Lines~6--12), where $W_{n,m} = 1$ if $|A{-}L|_{n,m} > \xi\sigma$, where $\sigma$ is the standard deviation of $A{-}L$ \textcolor{20170927}{and $\xi$ is a parameter to tune the threshold}. In Lines~13--15, the weighting matrix is converted to the 3-dimensional image cube $Z$, which has the same shape as the original data cube. The $m$th image in $Z$ is indicated by $Z_{*,*,m}$. Each image of $Z$ is processed as follows; Small masks, which are a size\footnote{The size of a mask is defined by the number of connected masked pixels.} smaller than $k$, are rejected in Line~17, and the remaining masks are dilated by a disk of radius $R$ (Line~18). Then, the image cube $Z$ is converted again into $W$ (Lines~20--22). This procedure is iterated until the remaining component $\sum_{n,m}W_{n,m}\left|A_{n,m}{-}L_{n,m}\right|^2$ becomes smaller than the given threshold $\epsilon$. \textcolor{20170927}{Finally, the background subtracted data cube $\hat{I}_{i,j,m}$ is derived from the subtraction $A-L$ (Lines~24--26). The data cube $\hat{I}_{i,j,m}$ is the movie data where the object is moving across the field-of-view. To obtain a combined image, a shift-and-add operation should be applied on $\hat{I}_{i,j,m}$. Note that the data cube $\hat{I}_{i,j,m}$ mainly consists of $f_{i,j}{\mathcal S}_{i,j,m}$. Thus, each frame of $\hat{I}_{i,j,m}$ should be divided by a flat frame, if available.}
\begin{table*}[p]
\caption{Slow-Scanning Decomposition}\label{algo:maskedgodec}
\centering
\begin{tabular}{rp{.8\linewidth}}
\hline\hline
\multicolumn{2}{l}{\textbf{Input:} $I_{i,j,m},r,\xi,\epsilon,k,R$} \\
\multicolumn{2}{l}{\textbf{Output:} $\hat{I}_{i,j,m}$} \\
\multicolumn{2}{l}{\textbf{Local Variables:}
$A,L,W \in \mathbb{R}^{N_XN_Y}{\times}\mathbb{R}^M$,
$Z \in \mathbb{R}^{N_X}{\times}\mathbb{R}^{N_Y}{\times}\mathbb{R}^M$} \\
\multicolumn{2}{l}{\textbf{Initialize:}
$A \leftarrow \mathbf{0}$, $L \leftarrow \mathbf{0}$,
$W \leftarrow \mathbf{0}$, $Z \leftarrow \mathbf{0}$} \\
1 &\textbf{for}
$i \leftarrow 1$ \textbf{to} $N_X$
and $j \leftarrow 1$ \textbf{to} $N_Y$
and $m \leftarrow 1$ \textbf{to} $M$\textbf{:} \\
2 &\quad $A_{i+N_Y(j-1),m} \leftarrow I_{i,j,m}$ \\
3 &\textbf{end for} \\
4 &\textbf{While}
{$\sum_{n,m}W_{n,m}|A{-}L|_{n,m}^2 > \epsilon$} \textbf{do} \\
5 &\quad $L \leftarrow {\rm BIRSVD}(A,W,r)^{\dagger 1}$ \\
6 &\quad \textbf{for}
$n \leftarrow 1$ \textbf{to} $N_X{\times}N_Y$
and $m \leftarrow 1$ \textbf{to} $M$\textbf{:} \\
7 &\qquad \textbf{if}
$\left|A-L\right|_{n,m} > \xi\sigma$ \textbf{then:}\\
8 &\quad\qquad $W_{n,m} \leftarrow 1$ \\
9 &\qquad \textbf{else:} \\
10&\quad\qquad $W_{n,m} \leftarrow 0$ \\
11&\qquad \textbf{end if} \\
12&\quad \textbf{end for} \\
13&\quad \textbf{for}
$i \leftarrow 1$ \textbf{to} $N_X$
and $j \leftarrow 1$ \textbf{to} $N_Y$
and $m \leftarrow 1$ \textbf{to} $M$\textbf{:} \\
14&\qquad $Z_{i,j,m} \leftarrow W_{i+N_Y(j-1),m}$ \\
15&\quad \textbf{end for} \\
16&\quad \textbf{for} $m\leftarrow1$ \textbf{to} $M$\textbf{:}\\
17&\qquad
$Z_{*,*,m} \leftarrow {\rm DropSmallMask}(Z_{*,*,m},k)^{\dagger 2}$ \\
18&\qquad
$Z_{*,*,m} \leftarrow {\rm BinaryDilation}(Z_{*,*,m},R)^{\dagger 3}$ \\
19&\quad \textbf{end for} \\
20&\quad \textbf{for}
$i \leftarrow 1$ \textbf{to} $N_X$
and $j \leftarrow 1$ \textbf{to} $N_Y$
and $m \leftarrow 1$ \textbf{to} $M$\textbf{:} \\
21&\qquad $W_{i+N_Y(j-1),m} \leftarrow Z_{i,j,m}$ \\
22&\quad \textbf{end for} \\
23&\textbf{end while} \\
24&\textbf{for}
$i \leftarrow 1$ \textbf{to} $N_X$
and $j \leftarrow 1$ \textbf{to} $N_Y$
and $m \leftarrow 1$ \textbf{to} $M$\textbf{:} \\
25&\quad $\hat{I}_{i,j,m} \leftarrow (A-L)_{i+N_Y(j-1),m}$ \\
26&\textbf{end for} \\
\hline\hline
\multicolumn{2}{l}{$^{\dagger 1}$obtain an $r$-lank approximation of the matrix $A$ with the weighting matrix $W$.} \\
\multicolumn{2}{l}{$^{\dagger 2}$erase segments whose sizes are smaller than $k$ pixels in $Z_{*,*,m}$.} \\
\multicolumn{2}{l}{$^{\dagger 3}$apply a binary-dilation process on $Z_{*,*,m}$ with the radius of $R$.}
\end{tabular}
\end{table*}
\section{Performance Verification}
\label{sec:observation}
\subsection{Observations with \textit{Subaru}/COMICS}
\label{sec:comics}
Experimental observations were carried out on 2015-07-27 (UTC) to estimate the performance of the ``slow-scanning'' observation. We obtained the images of \objectname[IRAS 17192+1836]{IRAS~17192+1836}, a bright variable star in Hercules, with \textit{Subaru}/COMICS both in a conventional chopping observation and the ``slow-scanning'' observation. The brightness of the object at $10\,{\rm \mu m}$ is about 3\,Jy \citep{beichman_infrared_1988}, which is sufficiently bright for an imaging observation with \textit{Subaru}/COMICS. Target information is listed in Table~\ref{tab:target}. Table~\ref{tab:obslog} summarizes the observations.
\begin{table}
\centering
\caption{Target Information}
\label{tab:target}
\begin{tabular}{lr}
\hline\hline
Target Name
& IRAS~17192+1836 \\
Coordinates$^a$
&  (17:21:26.18, +18:33:19.9) \\
Magnitudes
& $(B,\, V) = (12.81,\, 11.80)^a$ \\
IRAS Flux$^b$
& $F_{12\,\rm{\mu m}} = 3.25\,{\rm Jy}$,
$F_{25\,\rm{\mu m}} = 1.74\,{\rm Jy}$ \\
\hline
\multicolumn{2}{p{0.9\linewidth}}{$^a$the coordinates are from \citet{hog_tycho-2_2000}; $^b$the fluxes are from \citet{beichman_infrared_1988}.} \\
\end{tabular}
\end{table}
In the chopping observation, the chopping throw angle was set to $10\arcsec$, the position angle of the chopping was set to $0\arcdeg$ (North-South), and the chopping cycle was set to $2.17\,\mathrm{s}$, \textcolor{20170927}{which has been conventionally used in the observation with \textit{Subaru}/COMICS to achieve good performance\footnote{\textcolor{20170927}{The chopping interval of $2.17\,\mathrm{s}$ is from the performance of the tip-tilt secondary mirror of the \textit{Subaru} telescope. The instability of the secondary mirror at a high frequency ($\gtrsim 1\,\mathrm{Hz}$) will cause a poor observation efficiency.}} \citep{kataza_comics:_2000,sako_improvements_2003}. \citet{okamoto_improved_2003} reported that nodding was not necessary for bright objects since the residual pattern in a chop-subtracted image was almost negligible. Thus, nodding was not performed in the observation.} The exposure time for a single beam was about 0.964\,s with an overhead of about 0.120\,s. The overhead time corresponds to the time required for beam-switching. A single FITS data cube was composed of 22 exposures. The sequence was repeated 16 times. In total, 286 exposures were obtained.
In the ``slow-scanning'' observation, the chopping throw angle was set to $0\arcdeg$. The images were continuously obtained with the same exposure time and the same overhead time\footnote{This overhead time for beam-switching is practically not required in the ``slow-scanning'' observation. The observational efficiency can be improved by about 11\% by eliminating the overhead time.} as the chopping observation. The scanning speed was about {$0.{\!\arcsec}063\mathrm{s^{-1}}$} and the position angle of the scan was $-90\arcdeg$ (East to West). \textcolor{20170927}{Although the scanning PA was set orthogonal to the chopping PA by chance, we are confident that the result was not affected.} A single FITS data cube was composed of 88 exposures. The sequence was repeated 5 times. While 440 exposures were obtained in total, the first 286 exposures were used for the performance verification.
\begin{table*}[t]
\centering
\caption{Observation Log}
\label{tab:obslog}
\begin{tabular}[t]{lrr}
\hline\hline
& \multicolumn{1}{l}{Chopping Observation}
& \multicolumn{1}{l}{``Slow-Scanning'' Observation} \\\hline
Date
& 2015-07-27 (UTC) & 2015-07-27 (UTC) \\
Observing Time
& 06:25--06:52 & 10:13--10:21 \\
Airmass
& 1.007--1.029 & 1.318--1.358 \\
Filter
& 11.7$\,{\rm \mu m}$ & 11.7$\,{\rm \mu m}$ \\
Chopping Throw
& $10\arcsec$ & $0\arcsec$ \\
Chopping PA
& $0\arcdeg$ & --- \\
Chopping Cycle
& $2.167\,$s & --- \\
Scanning Speed
& --- & \textcolor{20170927}{$0.{\!\arcsec}063\mathrm{s}^{-1}$} \\
Scanning PA
& --- & $-90\arcdeg$ \\
Exposure Time
& $0.964\,$s/frame & $0.964\,$s/frame \\
Overhead Time
& $0.120\,$s/frame & $0.120\,$s/frame \\
Total Exposures
& 286\,frames & 286\,frames \\
\hline
\end{tabular}
\end{table*}
\begin{figure*}[tp]
\centering
\plottwo{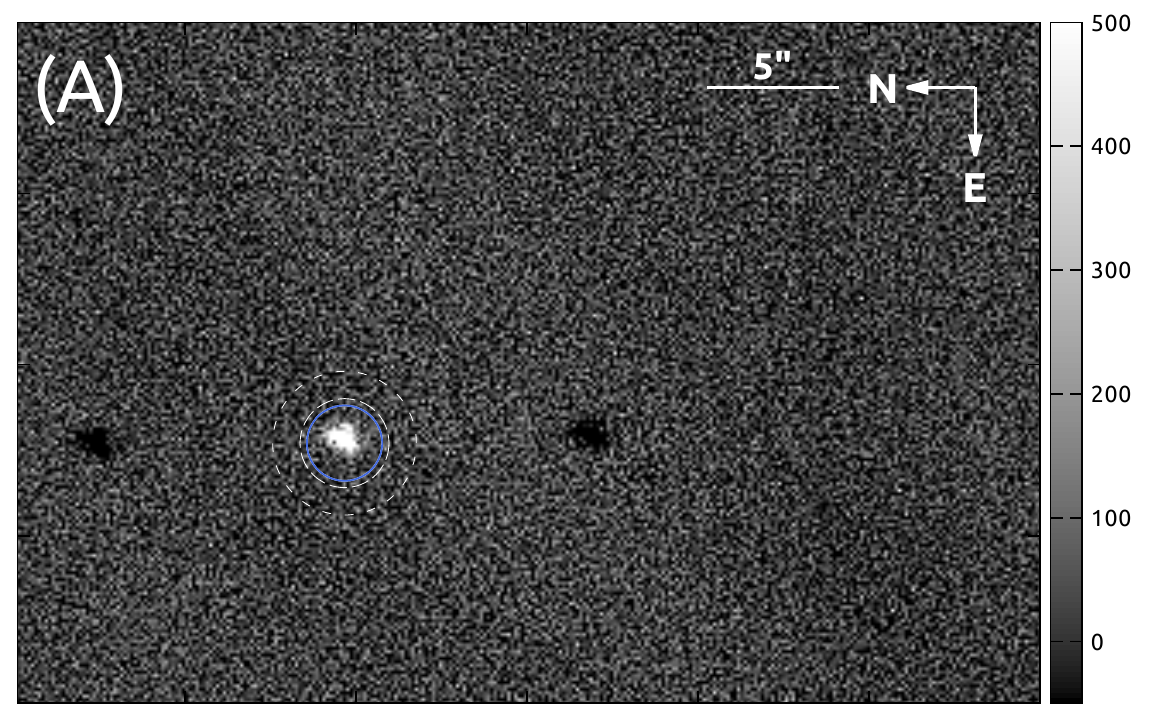}{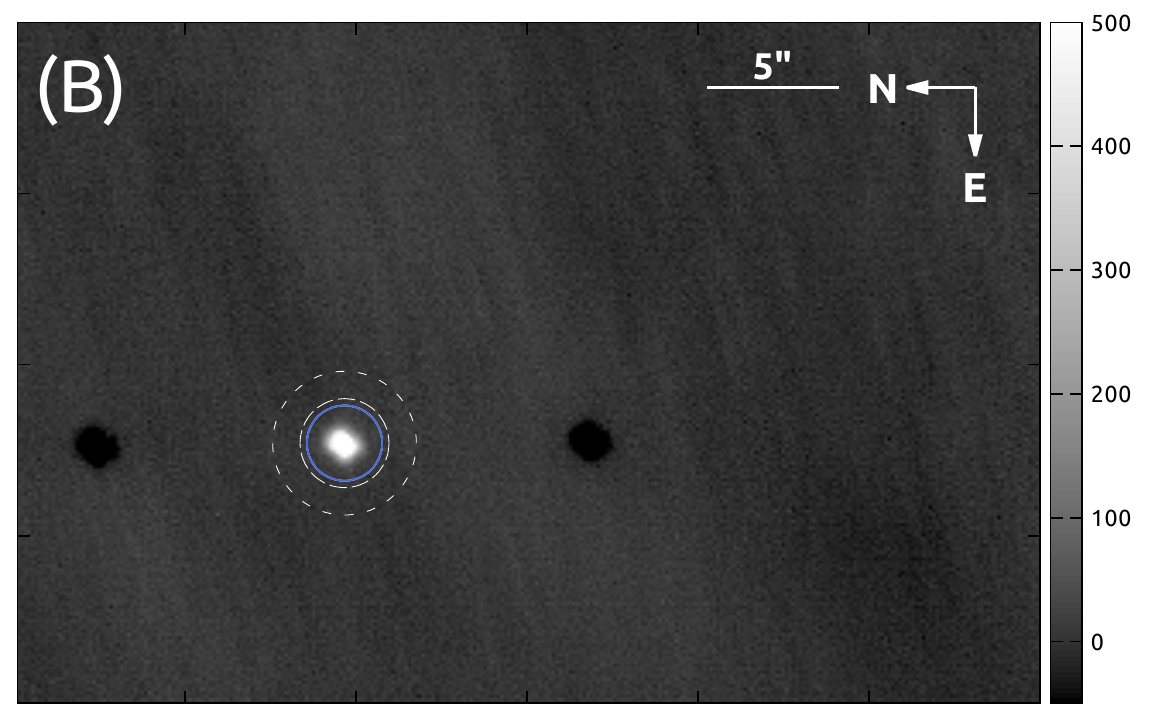}
\plottwo{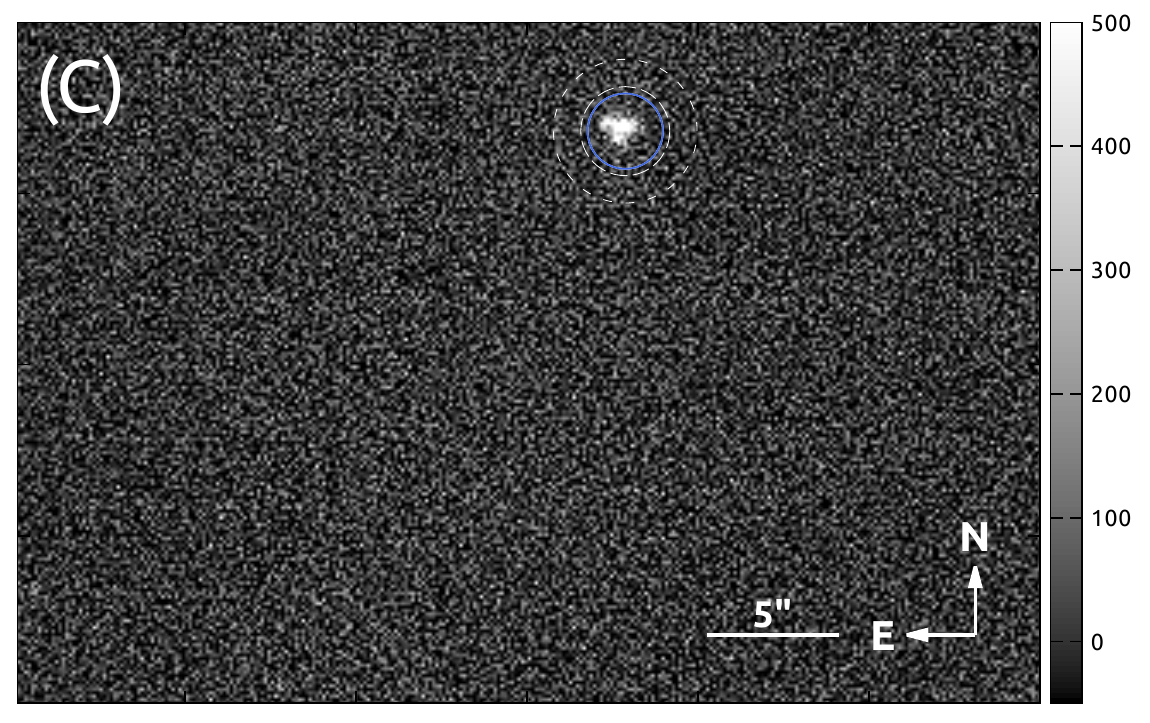}{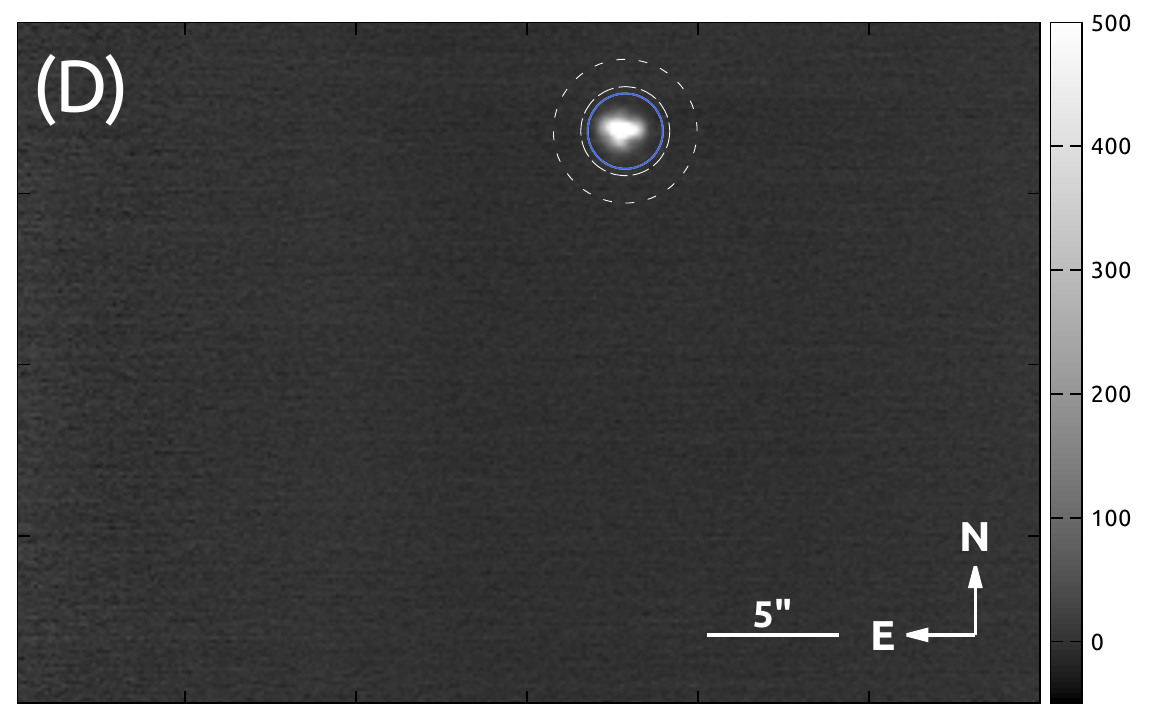}
\caption{Images obtained in the chopping and ``slow-scanning'' observations. Panel (A) shows one of the combined chopping pair images, while Panel (B) shows the combined image. Panel (C) shows a single exposure image from the non-low-rank FITS cube (\textit{see}, text), while Panel (D) illustrates the combined image in the ``slow-scanning'' observation.}
\label{fig:resultimages}
\end{figure*}
\begin{figure*}[tp]
\centering
\plottwo{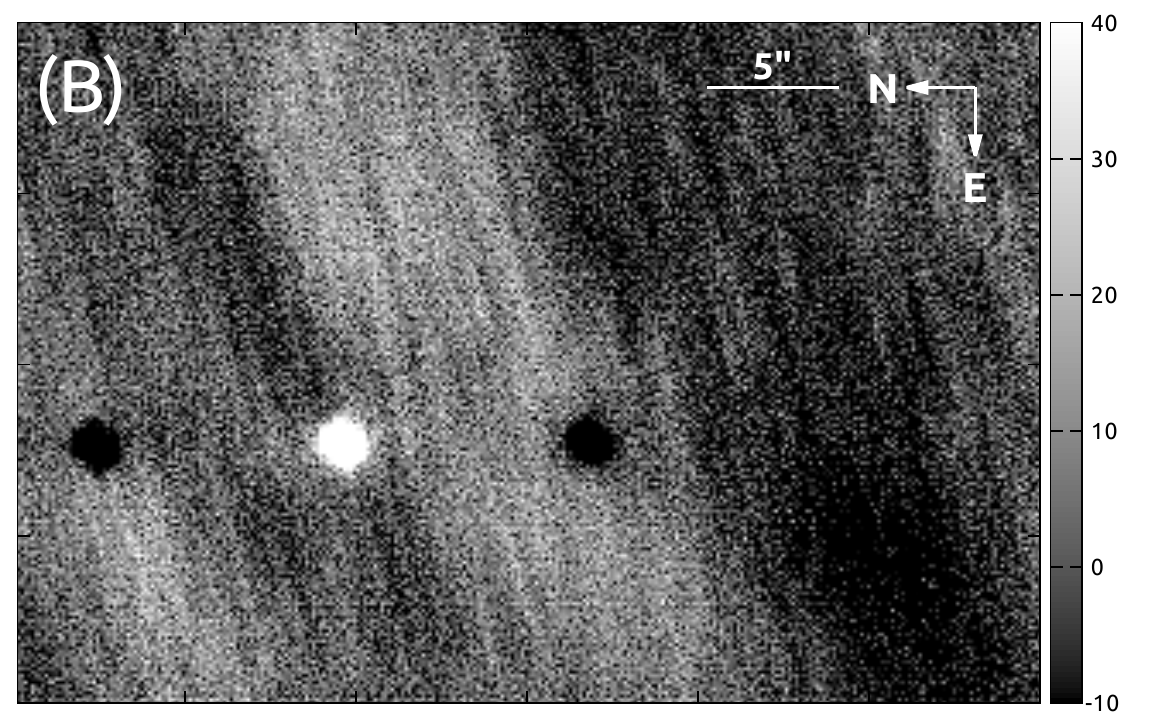}{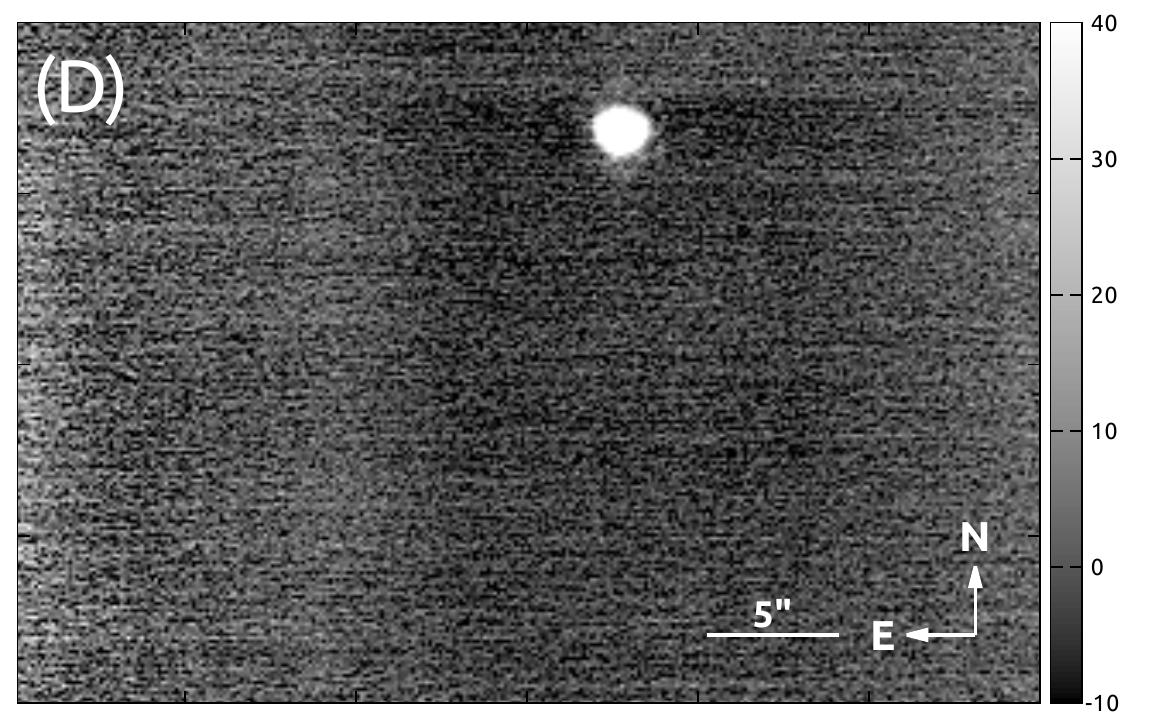}
\caption{Panels (B) and (D) of Figure~\ref{fig:resultimages} in different color scales.}
\label{fig:resultimages_highcontrast}
\end{figure*}
\begin{figure*}[tp]
\centering
\plotone{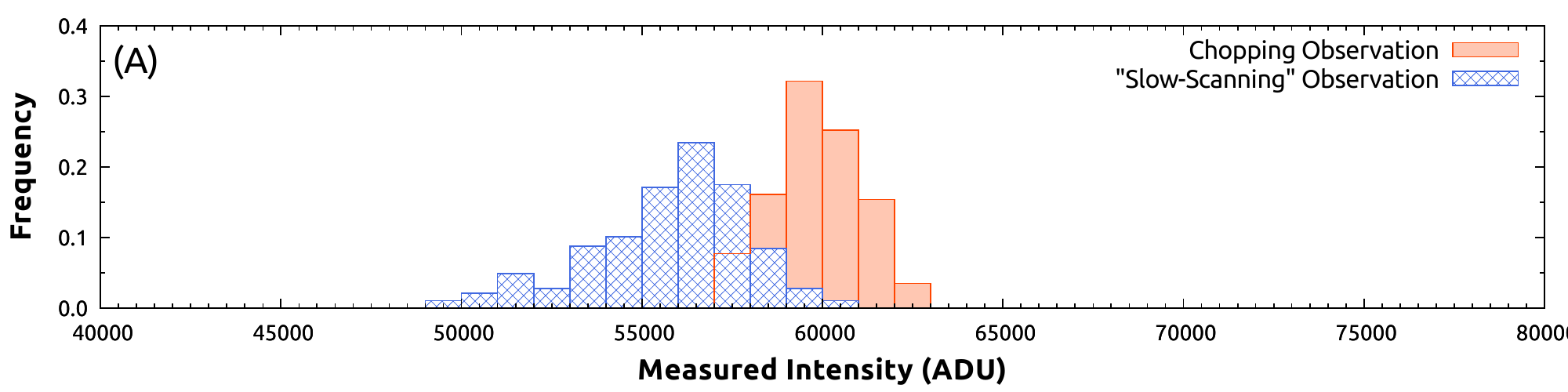}
\plotone{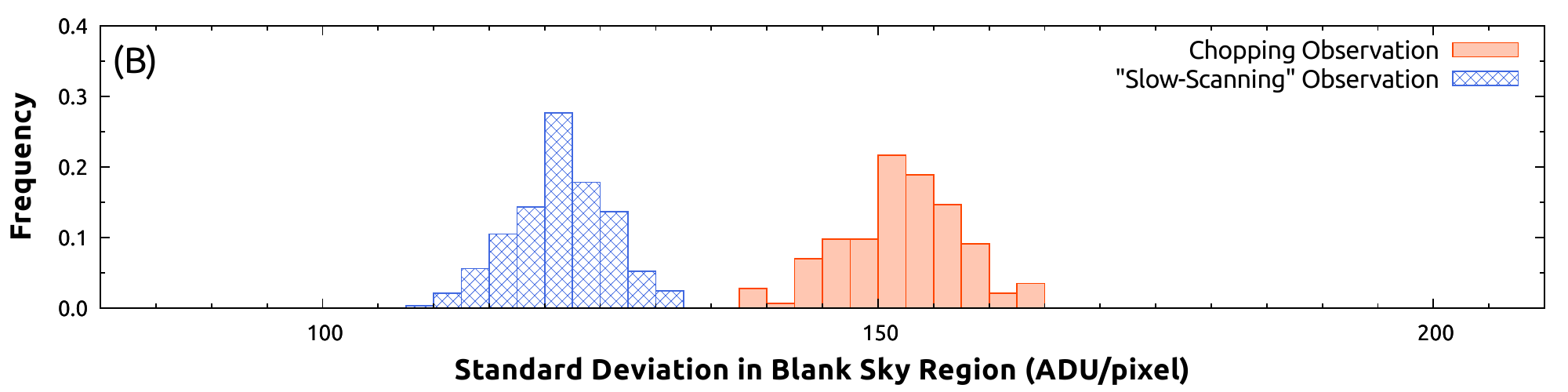}
\caption{Statistics of the photometric measurements for the chopping and ``slow-scanning'' observations. Panel (A) shows the distributions of the measured intensities, while Panel (B) shows those of the standard deviations in the blank sky region. The red solid bars show the results for the chopping observation. The blue cross-hatched bars indicate the results for the ``slow-scanning'' observation.}
\label{fig:singlestats}
\end{figure*}
\begin{table*}[tp]
\centering
\caption{Details of the Photometry on the Combined Images}
\label{tab:photresults}
\begin{tabular}{lrr}
\hline\hline
& \multicolumn{1}{c}{Chopping}
& \multicolumn{1}{c}{``Slow-Scanning''} \\\hline
Number of Stacked Frames
& 286 & 286 \\
Aperture Radius (pixel)
& 8 & 8 \\
Standard Deviation of Blank Sky Region (ADU)
& $11.05$ & $5.26$ \\
Integrated Intensity ($10^3$ADU)
& $59.34{\pm}0.16$ & $54.37{\pm}0.07$ \\
Signal-to-Noise Ratio
& 380 & 729 \\
\hline
\end{tabular}
\end{table*}
\subsection{Results of the Conventional Chopping Observation}
\label{sec:obs:chopping}
The data obtained in the chopping observation was composed of 286 exposures, or 143 chopping pairs. Subtracted images were derived from each chopping pair. \textcolor{20170927}{The COMICS system reads the detector current from all 16 channels at the same time. Uncertainty in the readout circuit causes the same noise pattern among the 16 channels (hereafter, the common-mode noise). The common-mode noise among output channels was subtracted using the channels where the object is not located \citep{sako_improvements_2003,okamoto_subaru_2012}.} Positive and negative sources in the subtracted images were combined by shift-and-add. No flat frame correction was applied, to keep the noise level from being affected by the uncertainty of a flat frame. Panel (A) of Figure~\ref{fig:resultimages} shows one of the combined chopping pair images. The intensity of the source was measured by aperture photometry using IRAF\footnote{IRAF is distributed by the National Optical Astronomy Observatories, which are operated by the Association of Universities for Research in Astronomy, Inc., under cooperative agreement with the National Science Foundation. distributed. IRAF is available in \url{http://iraf.noao.edu/}.}. \textcolor{20170927}{The blue solid lines indicate the photometric aperture size, while the annulus region indicated by the white dashed lines shows the blank sky region, where the background noise was evaluated.} The distributions of the intensities and the standard deviations of the background emission\footnote{The standard deviations were measured in the images after shift-and-add and multiplied by $\sqrt{2}$ to compare with those in the ``slow-scanning'' observation.} are illustrated in Figure~\ref{fig:singlestats}. The combined image was derived by averaging over 143 images, shown in Panel (B) of Figure~\ref{fig:resultimages}. Photometric results of the combined image are summarized in Table~\ref{tab:photresults}.
\subsection{Results of the ``Slow-Scanning'' Observation}
\label{sec:obs:slowscan}
First, we applied the same data reduction method on the data obtained in the ``slow-scanning'' observation and measured the standard deviation in a blank sky region. The standard deviation was almost the same as in the chopping observation. Thus we conclude that the conditions in the two observations were similar enough to be compared.
The non-redundant data cube was made from the 286 frames obtained during ``slow-scanning''. The adopted reduction parameters were as follows: $(r,\,\xi,\,k,\,R) = (12,\,3,\,3,\,4)$\footnote{In the \texttt{BIRSVD} calculation, a square of the second-order derivative with eight-order accuracy was selected as a regularization term. The strength of the regularization was set to 0.01. Refer to \citet{das_fast_2011} for details.}. The convergence of the algorithm was sufficiently fast, so that the iteration was terminated after 15 loops. No flat frame correction was applied. Panel (C) of Figure~\ref{fig:resultimages} shows one of the frames in the non-redundant data cube. The intensity of the source in each frame was measured by aperture photometry with the same parameters as in the chopping observation. \textcolor{20170927}{The blue solid circles and the white dashed lines are the same as in the chopping observation. The distributions of the measured intensities and the standard deviations in the blank sky regions are summarized in Figure~\ref{fig:singlestats}.} The 286 frames were combined by shift-and-add based on the motion of the telescope. The combined image is displayed in Panel (D) of Figure~\ref{fig:resultimages}. The intensity of the source in the combined image was measured by aperture photometry. The results are listed in Table~\ref{tab:photresults}.
\section{Discussion}
\label{sec:discussion}
\subsection{Comparison between the chopping and ``slow-scanning'' observations}
\label{sec:comparison_chop}
\subsubsection{Background Subtraction}
\label{sec:background_subtraction}
First, the residual patterns of blank sky regions in the combined images are compared. \textcolor{20170927}{In Figure~\ref{fig:resultimages_highcontrast}, Panels~(B) and (D) of Figure~\ref{fig:resultimages} are shown in different color scales to emphasize residual patterns.} The blank sky region in Panel~(B) of Figure~\ref{fig:resultimages_highcontrast} shows a slight but large-scale residual pattern, which is attributable to chopping: The instrumental emission pattern ($\mathcal{T}_{i,j}$) varies between the two chopping positions due to different light paths. This large-scale residual can be eliminated by the chop-and-nod observation. \textcolor{20170927}{The photometric results were not affected by the large-scale residual pattern, since the blank sky region in the photometry was chosen to be sufficiently small.} Panel~(D) of Figure~\ref{fig:resultimages_highcontrast} shows no large-scale residual background emission, since the light path is fixed during the ``slow-scanning'' observation. \textcolor{20170927}{Instead, horizontal stripe patterns, parallel to the scan direction, can be seen in the ``slow-scanning'' image. This pattern may be attributed to the shift-and-add process: the residual background emission has a particular pattern, and the stripe is created by dragging this residual pattern along the scanning direction. Similar patterns can be seen in \citet{heikamp_drift_2014}.} By comparing Panel (B) and Panel (D) of Figure~\ref{fig:resultimages_highcontrast}, we conclude that the image quality of the ``slow-scanning'' observation was as good as that of the conventional chopping observation in terms of background subtraction.
\subsubsection{Image Quality}
\label{sec:image_quality}
\begin{figure*}[tp]
\centering
\plotone{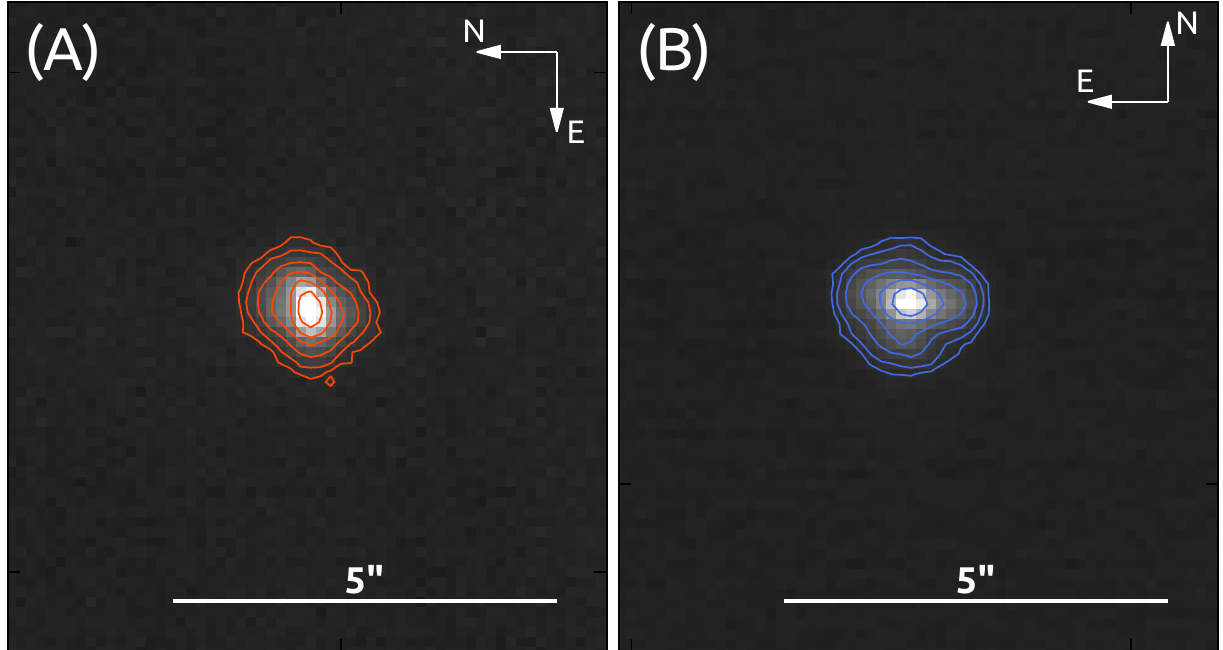}
\caption{Comparison of the point spread functions (PSF). Panels (A) and (B) show the close-up images obtained in the chopping and ``slow-scanning'' observations, respectively. The contours are drawn at the $5\sigma$, $10\sigma$, $20\sigma$, $40\sigma$, $80\sigma$, $160\sigma$, and $320\sigma$ levels, where $\sigma$ is the standard deviation of the blank sky regions.}
\label{fig:psfcomparison}
\end{figure*}

Figure~\ref{fig:psfcomparison} shows the close-up images of the object. Since the object is a point-like source, the image of the object is a good estimator of the point spread function (PSF). A distorted PSF implies poor image quality. Panel (A) of Figure~\ref{fig:psfcomparison} shows the image in the chopping observation, while Panel (B) of Figure~\ref{fig:psfcomparison} shows that of the ``slow-scanning'' observation. The contours are drawn at the $10\sigma$, $20\sigma$, $40\sigma$, $80\sigma$, $160\sigma$, and $320\sigma$, where $\sigma$ is the standard deviation in the blank sky regions.
\textcolor{20170927}{We measured the PSF size using IRAF. The MOFFAT FWHM, eccentricity, and position angle for the chopping observation were $0.{\!\arcsec}46$, $0.10$, and $-40\arcdeg$. Those for the ``slow-scanning'' observation were $0.{\!\arcsec}43$, $0.20$, and $85\arcdeg$. The PSF sizes suggest that the image quality in the ``slow-scanning'' observation was as good as in the chopping observation. The PSF in the ``slow-scanning'' observation was, however, slightly elongated along with the scan direction. The elongation was too large to be attributed to a motion blur during an exposure. The elongation was, in part, attributable to instability in the slewing speed of the \textit{Subaru} telescope in the non-sidereal tracking mode or in the positioning of the tip-tilt secondary mirror. Further investigation in different configurations is required to identify the origin of the elongation.}
\subsubsection{Photometric Measurements}
\label{sec:photmetric_measurements}
The photometric results of the chopping and ``slow-scanning'' observations are compared in Figure~\ref{fig:singlestats}. Panel (A) of Figure~\ref{fig:singlestats} suggests that the intensities obtained by the ``slow-scanning'' observation were about 7\% fainter than those obtained by the chopping observation. \textcolor{20170927}{The precipitable water vapor on 2015-07-27 was about 5\,mm and stable. The decrease is not attributable to the weather conditions.} \textcolor{20170927}{Since the flux was measured over many different pixels in the ``slow-scanning'' observation, the fluctuations in the flat frame were reduced. The fluctuation in the flat frame was about 4\% (\textit{see}, Appendix~\ref{sec:flatframe}), which was smaller than the decrease. Although the fluxes were measured in different positions between the chopping and ``slow-scanning'' observations, the decrease is not likely to be attributed to the flat frame. Since there was no plausible candidate to explain the decrease, we concluded} that some fraction of the flux was missed in the ``slow-scanning'' observation and the decrease, which could be attributable to the data reduction, was about 7\%.
Panel (B) of Figure~\ref{fig:singlestats} suggests that the standard deviation in the ``slow-scanning'' observation was about 20\% smaller than that in the chopping observation. In the chopping observation, noise power spectrum components around the chopping frequency are increased by a factor of ${\sim}2$. This is simply because science and reference frames have the same noise level (a detailed discussion is described in Appendix~\ref{sec:chopping}). On the other hand, a reference frame in the ``slow-scanning'' observation is generated from multiple exposures. The penalty of background subtraction in the ``slow-scanning'' observation is expected to be smaller than in the chopping observation. Consequently, the signal-to-noise ratios in the ``slow-scanning'' observation were higher than in the chopping observation \textcolor{20180301}{with \textit{Subaru}/COMICS}.
\begin{figure*}[tp]
\centering
\plotone{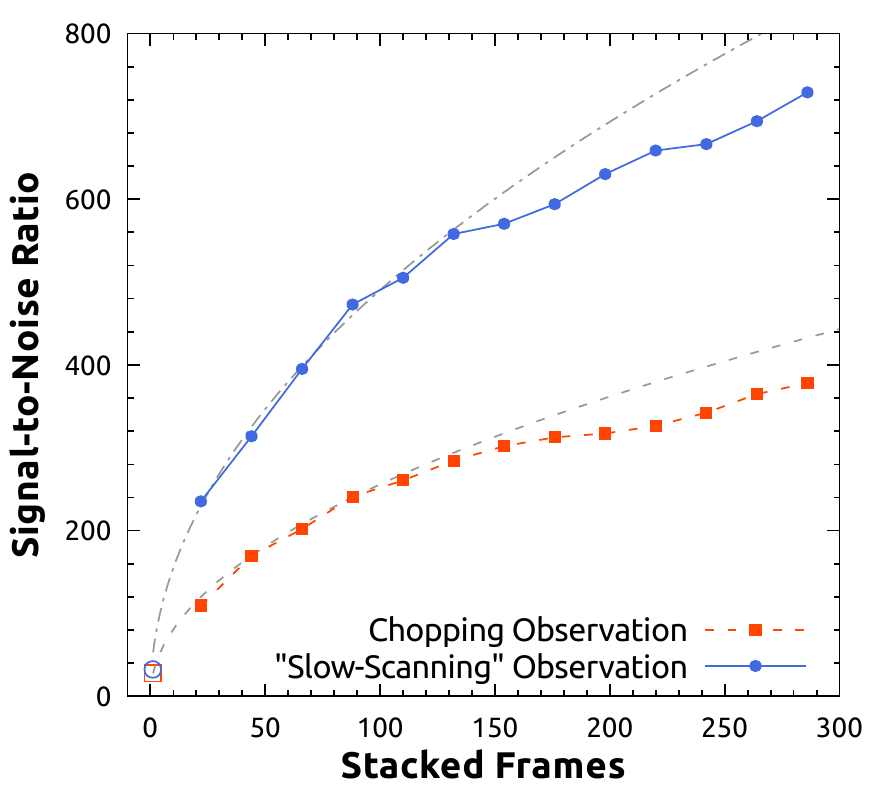}
\caption{Evolution of the signal-to-noise ratio. The red dashed line with the squares indicates the growth curve of the signal-to-noise ratios by stacking the frames in the chopping observation, while the blue solid line with the circles indicates those in the ``slow-scanning'' observation. The open symbols indicate the signal-to-noise ratios for single-frame photometry. The gray dashed and doted-dashed lines show evolution curves proportional to $\sqrt{N}$.}
\label{fig:snevo}
\end{figure*}
Table~\ref{tab:photresults} shows that the signal-to-noise ratio of the stacked image in the ``slow-scanning'' observation was about 88\% higher than in the chopping observation. The difference in the signal-to-noise ratios increased compared to the results of the single-frame photometry. We suppose that this is explained by a combination of two effects: \textcolor{20170927}{First}, since the object and the sky region were observed in different pixels in the ``slow-scanning'' observation, fluctuations in the flat frame were smoothed out. Second, residual noise was reduced since the subpixel shift-and-add procedure acted as a smoothing filter.
Figure~\ref{fig:snevo} illustrates how the signal-to-noise ratios improve by stacking the frames. \textcolor{20170927}{Since the two observations were carried out in the same detector settings, the horizontal axis approximately indicates the net observing times.} The dashed and doted-dashed lines are $\sqrt{N}$ curves for chopping and ``slow-scanning'' observations. The signal-to-noise ratios increased almost proportional to the $\sqrt{N}$ curves until ${\sim}150$ frames, indicating that the noise in the ``slow-scanning'' observation was not correlated among frames. Figure~\ref{fig:snevo} suggests that the signal-to-noise ratio in the ``slow-scanning'' observation increased more rapidly than in the chopping observation. The observational efficiency was improved in the ``slow-scanning'' observation \textcolor{20180301}{in the case of \textit{Subaru}/COMICS}.
\subsection{Comparison with other methods}
\label{sec:comparison_others}
The ``drift scanning'' method has been proposed as another method for the ground-based mid-infrared observation without chopping \citep{heikamp_drift_2014}. In the ``drift scanning'' observation, the field-of-view is moved much faster than in the ``slow-scanning'' observation, so that the object is obtained in different pixels before the atmospheric emission is changed. In the ``drift scanning'' method, the object is detected in different pixels and the background emission can be estimated using multiple exposures. Thus, the improvement of the observational efficiency is expected \citep{heikamp_drift_2014} as well as in the ``slow-scanning'' observation. The scanning speed in the ``drift scanning'' method $(v_\mathrm{drift})$ should be faster $\Theta \tau^{-1}$, where $\Theta$ is the angular size of the object and $\tau$ is the typical timescale of the variation in the atmospheric emission. Assuming that $\Theta = 1\arcsec$ and $\tau = 1\,$s, the required scanning speed is $\gtrsim 1\arcsec \mathrm{s}^{-1}$. To avoid the degradation of the PSF, the integration time of a single exposure should be shorter than $\theta v_\mathrm{drift}^{-1}$, where $\theta$ is the pixel scale of the detector. In a case where $\theta = 0.{\!\arcsec}13$ and $v_\mathrm{drift} = 1\arcsec\,\mathrm{s}^{-1}$, the integration time should be shorter than $0.13\,$s. In the case of \textit{Subaru}/COMICS, the full frame readout time is about 0.06\,s. The availability of the ``drift scanning'' method could be limited by the fastest readout time of the instrument.
The ``weighted average'' method is an extension of the dithering observation \citep{nakamura_method_2016}. The pointing is changed frequently, with an interval of about 15\,s, and the background emission is estimated by a linear combination of the obtained frames. By increasing the number of dithering positions, the observational efficiency in the ``weighted average'' method is expected to be improved as well as in the ``slow-scanning'' observation. The ``weighted average'' method requires neither fast slewing of the telescope nor a quick readout. \citet{nakamura_method_2016} reported that the image obtained in the ``weighted average'' method was as good as that in the chopping observation. On the other hand, the observational efficiency in the ``weighted average'' method can be affected by the overhead time for the telescope pointing.
\textcolor{20170927}{
\subsection{Applicability of the ``slow-scanning'' observation}
\label{sec:applicability}
}
There are several drawbacks in the ``slow-scanning'' observation. In the data reduction, we assume that the flat frame $f$ is constant in time. If the flat frame changes significantly during observations, the algorithm will fail to subtract background emission. The present results are based on the fact that the short-timescale variation in the flat frame was negligible compared to the pixel-to-pixel fluctuation in the flat frame (\textit{see}, Appendix~\ref{sec:flatframe}). \textcolor{20180301}{Although the stability of the flat frame was tested for \textit{Subaru}/COMICS, it is not confirmed that the short-timescale variation in the flat frame is negligible in observations with other facilities.} The ``slow-scanning'' observation will not be available for systems where the short-timescale variability in the flat frame is significant. In such situations, the conventional chopping observation or the ``drift scanning'' method should be used.
The next-generation mid-infrared instruments, MICHI and METIS \citep{packham_key_2012,brandl_status_2016}, are expected to use mid-infrared adaptive optics. Since the telescope is moved continuously, the ``slow-scanning'' observation is not compatible with adaptive optics. This significantly limits the applicability of the ``slow-scanning'' observation.
The data reduction in the ``slow-scanning'' observation is more time-consuming than in the chopping observation and the other methods mentioned above. Thus, the ``slow-scanning'' observation is not suitable for a quick look at an object. An alternative simple quick-look method should be developed.
A number of exposures are required to robustly extract non-redundant components. Trial image reconstructions with less than 64 exposures were not successful. This may put a lower limit on the required observing time, and thus reduce the observational efficiency for bright objects.
It is not confirmed that the ``slow-scanning'' observation works well for diffuse or faint objects. Further on-sky experimental observations are required.
Despite these difficulties, the ``slow-scanning'' observation can be easily performed with current telescopes and instruments and its efficiency is scarcely affected by overhead time for telescope movements. \textcolor{20180301}{Although the applicability of the ``slow-scanning'' observation was tested only for \textit{Subaru}/COMICS and for a point-source object, the observing method and algorithm we propose are potentially beneficial to other ground-based mid-infrared facilities.}
\section{Conclusion}
\label{sec:conclusion}
A new observational method for ground-based mid-infrared astronomy without chopping, the ``slow-scanning'' observation, is proposed. Images are continuously obtained as a three-dimensional data cube, while the field-of-view is slowly shifted. Non-redundant components, which mainly consist of signals from an astronomical object, are extracted from the obtained data cube by the \texttt{mGoDec} procedure. {The extracted frames are combined by shift-and-add. Finally, the combined image is obtained. The integration time of the combined image corresponds to the scan duration.}
The performance of the ``slow-scanning'' observation was tested in the experimental observations with \textit{Subaru}/COMICS. The image quality of the ``slow-scanning'' observation was as good as that of the conventional chopping observation. The observational efficiency in the ``slow-scanning'' observation was higher than in the chopping observation. This is supposed to be a combination of these effects: A background frame is made from multiple exposures; fluctuations in a flat frame are smoothed out since the field-of-view is moving; residual noise is suppressed since the shift-and-add process acts as a smoothing filter. \textcolor{20180301}{The present result suggests that the ``slow-scanning'' observation is practically available as an alternative to the conventional chopping observation in the case of \textit{Subaru}/COMICS. Although the applicability of the ``slow-scanning'' observation remains to be tested for other instruments, the method and algorithm we propose are potentially beneficial to other ground-based mid-infrared instruments.}
\acknowledgments
We thank Takuya Yamashita for detailed information on the tip-tilt secondary mirror of the \textit{Subaru} telescope and the COMICS operations. This research is supported by CREST and PRESTO, Japan Science and Technology Agency (JST), and in part, by JSPS Grants-in-Aid for Scientific Research (KAKENHI) Grant Number JP25120008, JP25103502, JP26247074, JP24103001, JP16H02158, and JP16H06341. This work was achieved using the grant of Joint Development Research by the Research Coordination Committee, National Astronomical Observatory of Japan (NAOJ).
\facility{Subaru (COMICS)}
\software{IRAF \citep{tody_iraf_1993,tody_iraf_1986}}

\appendix
\section{Short-Timescale Stability of the Flat Frame}
\label{sec:flatframe}
\begin{figure}
\centering
\plotone{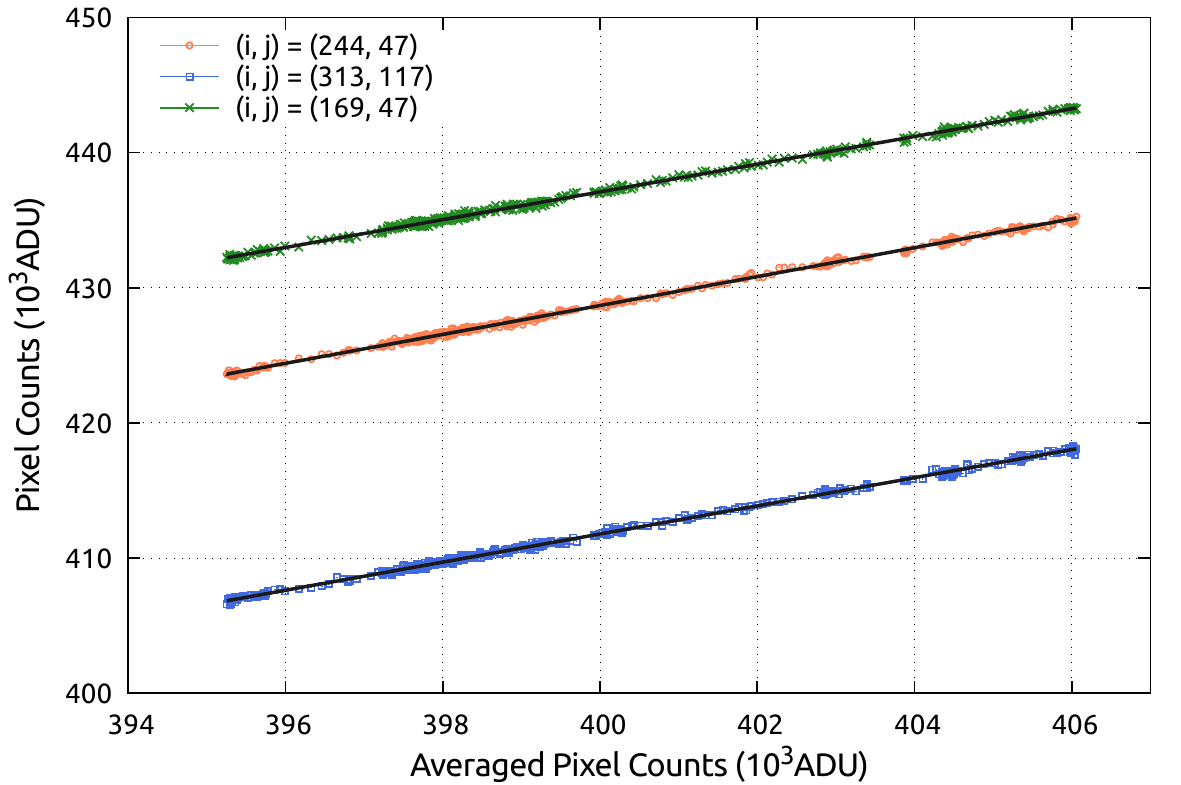}
\caption{Pixel counts plotted against the count averaged over the field-of-view. Statistical errors are as small as the symbol sizes. The black solid lines indicate the fitted curves (\textit{see}, text).}
\label{fig:flatframe}
\end{figure}

\textcolor{20170927}{In the proposed algorithm, we assume that the flat frame is stable in a short-time scale. The stability of the flat frame was evaluated based on the data we obtained. We used 440 exposures obtained with \textit{Subaru}/COMICS. First, we derived the time-sequence of the intensity averaged over the field-of-view.}
\begin{equation}
\label{eq:timeseq}
\textcolor{20170927}{\bar{I}_m = \frac{1}{N_XN_Y}\sum_{i,j}I_{i,j,m}.}
\end{equation}
\textcolor{20170927}{In Figure \ref{fig:flatframe}, the intensities of individual pixels are plotted against $\bar{I}_m$. The distributions of the data points are well-approximated by a linear function, the slope of which corresponds to the pixel response. We derived the slope for each pixel by fitting. The fitted curves are shown by the black solid lines in Figure~\ref{fig:flatframe}. The standard deviation of the slopes, which indicates a typical pixel-to-pixel fluctuation in the flat frame, was about $4.1\%$.}
\textcolor{20170927}{Figure~\ref{fig:flatframe} shows deviations from the fitting. The deviations were about 4 times larger than those expected from the photon Poisson noise. Part of the deviations should be attributed to the short-timescale variability in the flat frame. We assume that all the deviations originate from the short-timescale variability in the flat frame. The fractional deviation from the fit is derived by}
\begin{equation}
\label{eq:deviation}
\textcolor{20170927}{
\delta_{i,j,m} =
\frac{I_{i,j,m} - I^\mathrm{fit}_{i,j,m}}{I^\mathrm{fit}_{i,j,m}},}
\end{equation}
\textcolor{20170927}{where $I^\mathrm{fit}_{i,j,m}$ is the value derived by the fitting at $(i,\,j,\,t_m)$. The standard deviation of $\delta_{i,j,m}$, which indicates an upper limit on the short-timescale variation in the flat frame, was about $0.037\%$. Thus, we conclude that the flat frame was approximately constant in time during our observations.}
\section{Efficiency in Chopping Observation}
\label{sec:chopping}
\begin{figure*}
\centering
\plotone{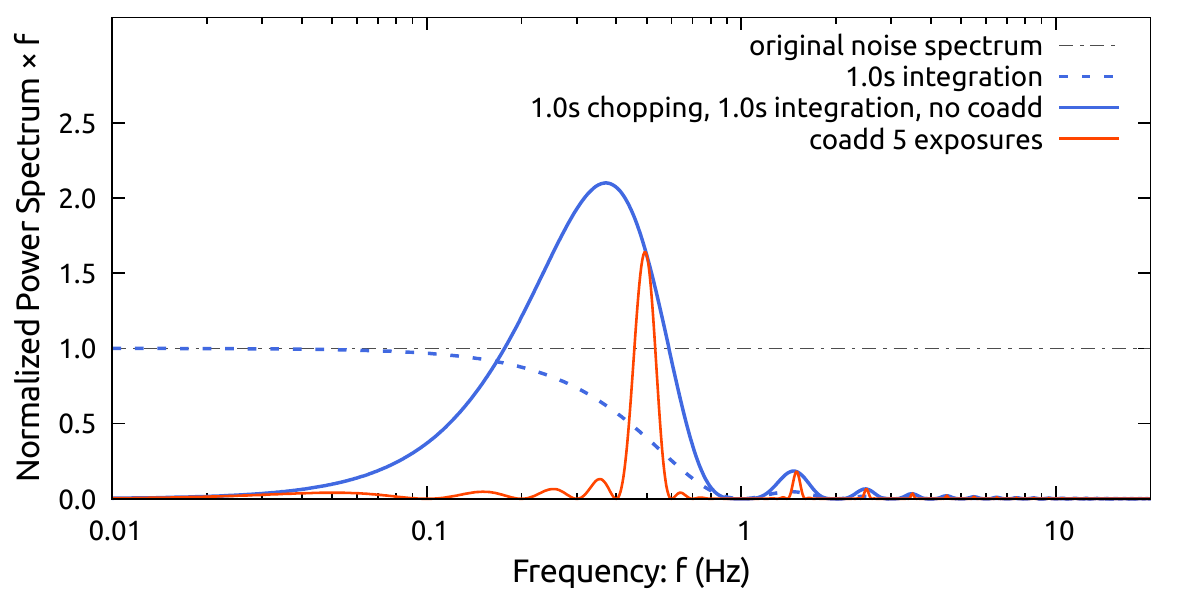}
\caption{Normalized noise power spectra in the case where ${\rm NPS}_{\mathcal A}(f) \propto f^{-1}$ (\textit{see}, text). The chopping-time separation $\delta{t}$ is set $1.0\,$s (0.5\,Hz chopping). The dashed lines show the power spectra for a single exposure before the subtraction of chopping pairs. The blue solid lines indicate the power spectra after the differentiation of chopping pairs. The red solid line shows the noise power spectrum after combining 5 chopping pairs.}
\label{fig:nps}
\end{figure*}

In the conventional chopping observation, we obtain a pair of images in different positions A and B, with a time separation of ${\delta}t$. By differentiating the pair, the signal from astronomical objects is extracted.
\begin{equation}
\label{eq:chopping}
I_{i,j}^{\rm A}(t) {-} I_{i,j}^{\rm B}(t{+}{\delta}t)
= f_{i,j} \bigl(
{\mathcal S}^{\rm A}_{i,j}{-}{\mathcal S}^{\rm B}_{i,j}
{+}{\mathcal A}_{i,j}(t){-}{\mathcal A}_{i,j}(t{+}{\delta}t)
\bigr),
\label{eq:chopping_approx}
\end{equation}
where the differences in emission from the instrument and the telescope between the chopping positions is neglected for the sake of simplicity. In a limit of ${\delta}t \rightarrow 0$, atmospheric emission is perfectly canceled out. Practically, ${\delta}t$ should be a short but finite value. The time separation required to eliminate the variation in the atmospheric emission depends on the power spectrum of ${\mathcal A}(t)$. Suppose that the Fourier transformation of ${\mathcal A}(t)$ is $\tilde{\mathcal A}(f)$, the noise power spectrum (NPS) of ${\mathcal A}(t)$ is described by
\begin{equation}
\label{eq:esd_A}
{\rm NPS}_{\mathcal A}(f)
= \tilde{\mathcal A}(f)\tilde{\mathcal A}^\dagger(f),
\end{equation}
where $X^\dagger$ is the complex-conjugate of $X$. In practice, the observed atmospheric signals are integrated over an exposure time and averaged over exposures. Taking into account the integration time of $\Delta{t}\,({<}\delta{t})$ and the number of exposures $N$, an effective atmospheric emission component is given by the convolution of ${\mathcal A}(t)$ and a window function:
\begin{equation}
\label{eq:eff_A}
{\mathcal A^*}(t) = \int_{-\infty}^{\infty}
{\mathcal A}(t')\frac{1}{N}\sum_{n=0}^{N-1}
{\rm rect}\left(\frac{t-2n{\delta}t-t'}{\Delta{t}}\right)
{\rm d}t',
\end{equation}
where ${\rm rect}(x)$ is a rectangular function. Equation~(\ref{eq:esd_A}) describes the noise energy in the frequency range between $f$ and $f+{\rm d}f$. The differentiation of ${\mathcal A^*}(t)$ with the time separation of ${\delta}t$ is defined by ${\alpha}(t,{\delta}t)$:
\begin{equation}
\label{eq:alpha}
{\alpha}(t,{\delta}t) = {\mathcal A^*}(t) - {\mathcal A^*}(t+{\delta}t).
\end{equation}
The energy spectral distribution of ${\alpha}(t,{\delta}t)$ is given as:
\begin{eqnarray}
{\rm NPS}_{\alpha}(f \,|\, {\delta}t,{\Delta}t, N)
& \propto
& {\rm NPS}_{\mathcal A}(f)
\,\,2\bigl(1 - \cos 2\pi f\,{\delta}t\bigr)
\Bigl|{\rm sinc} \left(f\Delta{t}\right) \Bigr|^2
\left|
\frac{1}{N}\sum_{n=0}^{N-1}{\rm e}^{4{\pi}in{\delta}tf}
\right|^2, \nonumber\\
& =
& {\rm NPS}_{\mathcal A}(f)
\cdot\mathcal{F}_1(f\delta{t})
\cdot\mathcal{F}_2(f\Delta{t})
\cdot\mathcal{F}_3(N,f\delta{t}),
\label{eq:esd_alpha}\\
&& \text{where}~
\left\{~~\begin{array}{l}
\mathcal{F}_1(f\delta{t})
= 2\bigl(1 - \cos 2\pi f\,{\delta}t\bigr),\\
\mathcal{F}_2(f\Delta{t})
= \Bigl|{\rm sinc} \left(f\Delta{t}\right) \Bigr|^2,\\
\mathcal{F}_3(N,f\delta{t})
= \left|\frac{1}{N}\sum_{n=0}^{N-1}{\rm e}^{4{\pi}in{\delta}tf}\right|^2
\end{array}\right.
\end{eqnarray}
Equation~(\ref{eq:esd_alpha}) indicates that the chopping procedure is recognized as a product of three filtering functions. The filter $\mathcal{F}_1$, a differentiation filter, works as a high-pass filter. As long as $2\pi f\,{\delta}t \simeq 0$, noise components are suppressed. Instead high-frequency components are increased by this filter. The filter $\mathcal{F}_2$, an integration filter, works as a low-pass filter. Noise components at higher than about $\Delta{t}^{-1}\,$Hz are suppressed. The filter $\mathcal{F}_3$ is an averaging filter, which ensures that the integral of ${\rm NPS}_{\alpha}$ over $f$ is almost proportional to $N^{-1}$. The averaging filter does nothing when $N=1$, but it suppress the noise components between $(2N{\delta}t)^{-1}$ and $(N-1)(2N{\delta}t)^{-1}$\,Hz.
Figure~\ref{fig:nps} illustrates how the chopping works. We assume that ${\rm NPS}_{\mathcal A}(f) \propto f^{-1}$. High-frequency noise components are suppressed by integration. The blue dashed line indicates the noise spectrum after 1.0\,s integration. High-frequency noise components are suppressed. The solid line indicates the NPS after the differentiation of 0.5\,Hz chopping pairs ($\delta{t} = 1.0\,\mathrm{s}$). The noise components below about 0.2\,Hz are suppressed by chopping, while noise components around 0.5\,Hz are increased by a factor of around 2. The red solid line indicates the NPS after stacking 5 chopping pairs. Low-frequency noise components are well-suppressed by stacking.
\end{document}